\documentclass[twocolumn,prl,showpacs]{revtex4}
\usepackage{graphicx}

\usepackage{dcolumn}

\usepackage{tabularx}

\usepackage{float}

\usepackage{amsmath}

\newcommand{\comment}[1]{}

\begin{document}


\title{Optical conductivity of rattling phonons in type-I clathrate Ba$_8$Ga$_{16}$Ge$_{30}$} 
\author{T. Mori, S. Goshima, K. Iwamoto, S. Kushibiki, H. Matsumoto, and N. Toyota} 
\affiliation{Physics Department, Graduate School of Science, Tohoku University, Sendai 980-8578, Japan}
\author{K. Suekuni$^1$, M. A. Avila$^1$, and T. Takabatake$^{1,2}$}
\affiliation{$^1$Department of Quantum Matter, ADSM, $^2$Institute for Advanced Material Research, Hiroshima University, Higashi-Hiroshima 739-8530, Japan}
\author{T. Hasegawa, N. Ogita, and M. Udagawa} 
\affiliation{Graduate School of Integrated Arts and Sciences, Hiroshima University, Higashi-Hiroshima, 739-8521, Japan}

\date{\today}
\begin{abstract}
A series of infrared-active optical phonons have been detected in type-I clathrate Ba$_8$Ga$_{16}$Ge$_{30}$ by terahertz time-domain spectroscopy. The conductivity spectra with the lowest-lying peaks at 1.15 and 1.80\,THz are identified with so-called rattling phonons, i.e., optical modes of the guest ion Ba$^{2+}(2)$ with $T_{1u}$ symmetry in the oversized tetrakaidecahedral cage. The temperature dependence of the spectra from these modes are totally consistent with calculations based on a one-dimensional anharmonic potential model that, with decreasing temperature, the shape becomes asymmetrically sharp associated with a softening for the weight to shift to lower frequency. These temperature dependences are determined, without any interaction effects, by the Bose-factor for optical excitations of anharmonic phonons with the nonequally spaced energy levels.
\end{abstract}


\maketitle

During the past decade, thermoelectric materials such as clathrates and filled-skutterudites have renewed an interest in phonons\cite{JPSJ2008}. These conductors are formed by polyhedral building blocks, where each polyhedral cage accommodates a guest ion like an alkali-earth or rare-earth element. If the cage is oversized, the guest ions can vibrate with large amplitude around an on-center or off-center site in the cage potential. These Einstein-like local modes have been called \textit{rattling phonons}. The renewed interest above has been paid to interactions of these rattling phonons with acoustic phonons propagating through the cage network and carrying heat entropy, and, equally or more interestingly, with conduction electrons. However, these issues are still far from being well understood that even the charge dynamics of a single ion in rattling vibration have been little known. 

This paper reports on the optical conductivity of rattling phonons detected in a type-I clathrate Ba$_8$Ga$_{16}$Ge$_{30}$ (hereafter abbreviated as BGG), featuring the anharmonicity effects on the conductivity spectra in comparison with model calculations.
 
This compound belongs to a family of ternary type-I clathrates having cubic crystal structure with space group $Pm\bar{3}n$\cite{Kasper1965}. The unit cell of the host framework consists of 6 tetrakaidecahedra and 2 dodecahedra. The latter smaller cages around 2$a$ sites occupy the \textit{body-centered-cubic} sites, while the former oversized cages around 6$d$ sites line up along the principal axes by sharing both the pentagonal faces with the dodecahedral cages and the hexagonal faces with neighboring tetrakaidecahedra. Every cage encapsulates one Ba$^{2+}$ ion, and the guest ions in the smaller ($2a$) and oversized ($6d$) cages are defined as Ba$^{2+}$(1) and (2), respectively. These guest ions satisfy the Zintl rule for charge compensations that the more electro-positive guest atom donates electrons to the more electro-negative cage; 16 electrons are transferred from 8 Ba atoms to 16 Ga atoms of the cage. With finely tuned Ga/Ge concentration, therefore, the system can be a \textit{heavily} doped semiconductor, in general, having both the charge-carrier's sign and density controlled. 

So far the electrical resistivity, thermal conductivity and specific heat measurements have been made on BGG having different carrier densities of $n$- or $p$-type carriers\cite{Sales2001, Avila2006}. The nuclear density map for the Ba$^{2+}$(2) ion in the oversized cage obtained by neutron diffraction\cite{Sales2001} indicates that Ba$^{2+}$(2) resides on center, while other ions, Sr and Eu, move away to off-center sites, being supported by the band structure calculations\cite{Madsen2005}. On the contrary, recent more detailed structural analyses\cite{Christensen2006} indicate that the Ba$^{2+}$(2) ions are slightly displaced from the center depending on temperature and also carrier-type. This structural indication seems to be consistent with Raman scattering measurements of low lying additional modes\cite{Takasu2006}.  

\begin{figure}[h]
\includegraphics[width=8cm]{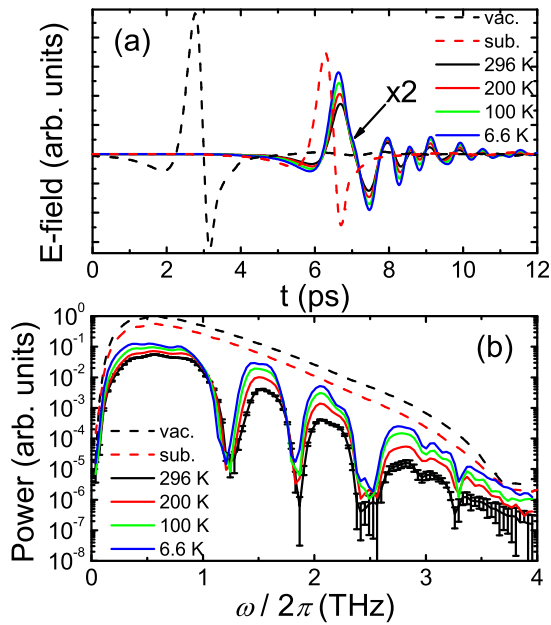}
\caption{(color online) (a) Time-evolution of THz-wave electric fields obtained from three different sets of measurements; vacuum, the substrate only and the type-I Ba$_8$Ga$_{16}$Ge$_{30}$ (BGG) sample glued onto the substrate and (b) the Fourier-transformed power spectra $|E(\omega)|^{2}$.}
\label{fig1}
\end{figure}
 
 Single crystals of $p$-type BGG are grown by a self-flux method\cite{Avila2006b}. To obtain transmitting signals as strong as possible through the conductive sample, single crystal disks of 2\,-\,5\,mm in diameter and 0.5\,-\,1.0\,mm in thickness are polished down to about 20\,$\mu$m in thickness using diamond abrasive films. Terahertz time-domain spectroscopy (THz-TDS) measurements covering the frequency range of 0.2 - 3.5\,THz (0.8 - 14.5\,meV) are carried out with the spectrometer (RT-20000, Tochigi Nikon Co. Ltd) which uses a standard technique for the transmission configuration, see, e.g., our preceding paper\cite{Mori2008}.

Figure \ref{fig1}(a) shows the time-evolution of transmitting electric fields through the vacuum, the substrate only and the sample glued onto the substrate, while Fig. \ref{fig1}(b) shows the Fourier-transformed power spectra. Then the refractive index spectra are determined by taking into account multiple reflections at the surface and boundary between the sample and substrate\cite{Mori2008}.

\begin{figure}
\includegraphics[width=8cm]{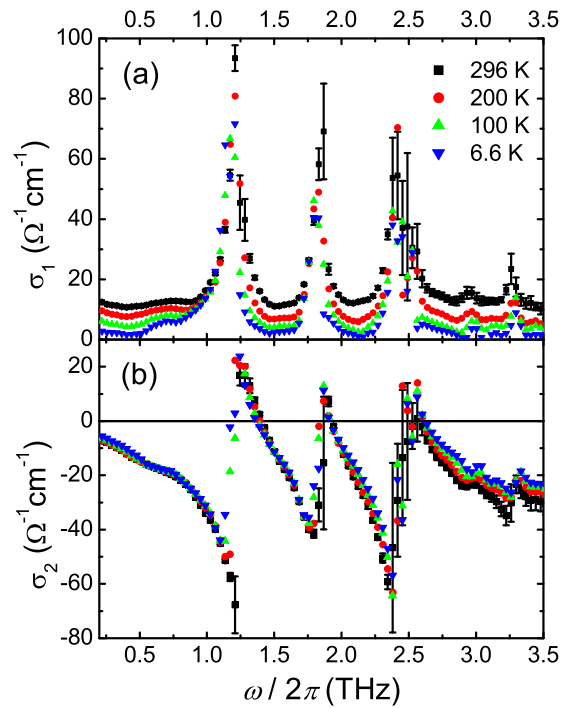}
\caption{(color online) Temperature-dependent conductivity spectra $\hat{\sigma}(\omega) = \sigma_{1}(\omega) + i \sigma_{2}(\omega)$ of BGG. For convenience, 1\,THz = 33\,cm$^{-1}$ = 4.1\,meV = 48\,K.}
\label{fig2}
\end{figure}

Figure \ref{fig2} shows the conductivity spectra $\hat{\sigma}(\omega) = \sigma_{1}(\omega) + i \sigma_{2}(\omega)$. The real part $\sigma_{1}(\omega)$ takes several distinct peaks, while the imaginary part $\sigma_{2}(\omega)$ shows correspondingly the derivative with frequency. These discrete spectra, reminiscent of Lorentz spectra for optical phonons, are superimposed on the monotonous background conductivity. The magnitude of the background at room temperature is about 10\,$\Omega^{-1}\textrm{cm}^{-1}$ comparable to the dc conductivity and, with decreasing temperature, decreases almost down to zero, again consistent with the temperature dependence of the dc conductivity\cite{Avila2006b}. Therefore it is concluded that the conductivity spectra consist of both an almost constant Drude-like contribution due to the semiconducting carriers and the Lorentz-like spectra from optical phonon modes.
\begin{table}
\caption{Calculated phonon frequencies and fitting parameters (6.6\,K) of BGG. Due to the broad spectra around 2.4\,THz (Fig.\,\ref{fig2}), the analysis for $\nu_{3}$ and $\nu_{4}$ is made with a single Lorentz curve (Eq.\,\ref{fitting}).}
\label{tab1}
\begin{center}
{\renewcommand\arraystretch{1.2}
\setlength{\arrayrulewidth}{0.2pt}
\begin{tabularx}{0.48\textwidth}{@{\extracolsep{\fill}}cccccc} \hline\hline
$label$ & $ atom$ & $calc.$ & $\omega_{0,i}$ & $\Gamma_{i}$ & $S_{i} \times 10^{-13}$ \\
& & $(\textrm{THz})$ & $(\textrm{THz})$ & $(\textrm{THz})$ & $(\Omega^{-1}\textrm{cm}^{-1}s^{-1})$ \\ \hline
$\nu_{1}$ & Ba(2) & 0.96 & 1.15 & 0.09 & 8.8 \\
$\nu_{2}$ & Ba(2) & 1.85 & 1.80 & 0.12 & 4.9 \\
$\nu_{3}$ & cage & 2.34 & \raisebox{-.7em}[0pt][0pt]{$\sim 2.4$} & \raisebox{-.7em}[0pt][0pt]{0.16} & \raisebox{-.7em}[0pt][0pt]{$\sim 6$} \\
$\nu_{4}$ & Ba(1) & 2.41 & & & \\
$\nu_{5}$ & cage & 2.70 & 2.98 & 0.06 & 0.4(1) \\
$\nu_{6}$ & cage & 3.05 & 3.30 & 0.07 & 0.6(4) \\ \hline\hline
\end{tabularx}}
\end{center}
\end{table}

For the assignment of the phonon modes, we have made first-principle calculations\cite{Takasu2006} based on the density functional approximation for infrared-active optical modes of $T_{1u}$ symmetry. Table \ref{tab1} lists the phonon frequencies labeled as $\nu_{i}(i = 1 - 6)$ together with fitting parameters at $T$ = 6.6\,K. The low-lying peaks observed at 1.15 and 1.80\,THz  can be assigned as the modes labeled $\nu_{1}$ and $\nu_{2}$ of the Ba$^{2+}$(2) in the oversized cage vibrating within the plane perpendicular to the fourfold axis and along the fourfold axis, respectively. It is noted that the Einstein temperature of about 60 K (=1.2\,THz) calculated from the atomic displacement parameters\cite{Christensen2006} is close to $\nu_1$. The broad peak around 2.4\,THz can be considered as an overlapped spectrum consisting of both the collective cage mode  $\nu_{3}$ and the Ba$^{2+}$(1) mode $\nu_4$ in the smaller cage. Higher frequencies of 2.98 and 3.30\,THz are rather close to the collective cage modes $\nu_{5}$ and $\nu_{6}$, respectively.
\begin{figure}
\includegraphics[width=8cm]{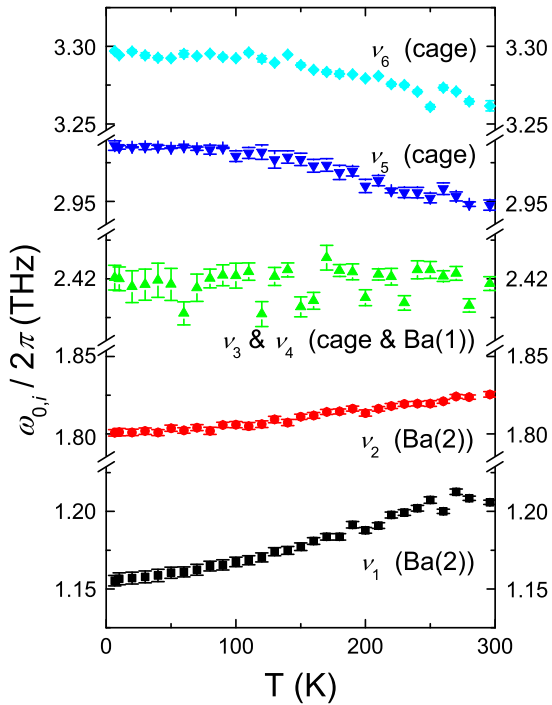}
\caption{(color online) Temperature dependence of peak frequencies of BGG for $\nu_{i} (i = 1 -6)$.}
\label{fig3}
\end{figure}

To extract each phonon contributions $\sigma_{1,i}(\omega)$ from the total spectra, we subtract the temperature-dependent Drude contributions by assuming a linear dispersion. Here we adjust the parameters of the linear curve in order to keep the sum rule that the spectral weight $S_{i}$ of $\nu_{i}$ is conserved irrespective of temperature;
\begin{equation}
\int_{0}^{\infty}\sigma_{1,i}(\omega)d\omega = S_{i}= \pi N_{i}q_{i}^{2}/2M_{i}, 
\label{sumrule}
\end{equation}
where $M_{i}$, $N_{i}$, and $q_{i}$ are the mass, density and charge of atoms involved in the $\nu_{i}$ phonon mode. Then to the phonon spectra from $\nu_{i}$ we apply the Lorentz-type conductivity written by
\begin{equation}
\sigma_{1,i}(\omega)=\frac{(2/\pi)S_{i}\omega^{2}\Gamma_{i}}{(\omega_{0,i}^2-\omega^2)^2+(\omega\Gamma_{i})^2},
\label{fitting}
\end{equation} 
where $\Gamma_{i}$ is the relaxation rate and $\omega_{0,i}$ is the resonant frequency. Here $S_{i}$ is analyzed from Eq.\ref{sumrule} using the lowest-temperature data having negligibly small Drude contribution, see Table \,\ref{tab1}. The observed ratio of $S_{1}/S_{2} = 1.8$ is close to the theoretical value of 2 taken into account of the degrees of freedom for $\nu_1$ and $\nu_2$ modes. Furthermore the total contribution ($S_{1} + S_{2}$) from these Ba(2)-ions' modes is about $13.7 \times 10^{13}$, which is larger by factor 4 than $3.4 \times 10^{13}$ calculated with $M = M_{\textrm{Ba}}$, $N = 6/V_{u}$ ($V_{u} = 10.76^{3}\,\textrm{\AA}^3$ is the unit cell volume\cite{Christensen2006}) and $q=2e$ for Ba(2) ions. Judging from the uncertainty in deriving the absolute value of $\sigma_1$, it is not clear at present whether this discrepancy is intrinsic or not. To note, however, it may be an open issue to study how the charge and mass of the guest ion are effectively modified when coupled to the encapsulating cage. For $\nu_5$ and $\nu_6$ modes of cages, the small spectral weight as listed in the table may be reasonable.

The peak frequencies $\omega_{0,i}/2\pi$ are plotted as a function of temperature in Fig.\,\ref{fig3}. With decreasing temperature down to 6.6\,K, the low-lying $\nu_1$ and $\nu_2$ modes soften by 4.2\%, being consistent with the Raman-active modes at 1.05\,THz (35\,cm$^{-1}$) with $E_g$ symmetry\cite{Takasu2006}, and 1.8\% respectively. On the other hand, other higher-frequency modes remain almost constant or harden slightly. In general the phonon hardening can be conventionally expected due to the lattice contraction at lower temperatures, so that the magnitude of the softening as observed in $\nu_1$ and $\nu_2$ might be underestimated due to an overcompensation for the hardening as observed in the cage modes $\nu_5$ and $\nu_6$. Thus the present studies reproduce the softening phenomena in rattling modes of guest ions, which have been so far interpreted to be an evident proof for an anharmonicity effect in these localized modes.  

In order to shed more light into the anharmonicity effect, we focus on the lowest-lying mode $\nu_1$ to compare with model calculations for the optical conductivity. We assume the following Hamiltonian describing the motion of a single ion in a one-dimensional anharmonic potential (1D-AP) including the quadratic and quartic terms for the displacement $x$,
\begin{equation}
H = \frac{p^{2}}{2M}+ \frac{1}{2}kx^{2}+\frac{1}{4}\lambda x^{4},
\label{H}
\end{equation}
where $p$ and $M$ are the momentum and mass (here, suppose Ba ions). The constants $k$ and $\lambda$ are assumed to be positive for a single well. Using a linear response theory, the optical conductivity $\hat{\sigma}(\omega)$ from the anharmonic phonon is expressed as
\begin{equation}
\begin{split}
\hat{\sigma}(\omega)/\sigma_{0} = i\omega\sum_{\omega_{nm}>0}|\langle n|x/x_{0}|m\rangle |^{2}\frac{e^{-\beta E_{m}}-e^{-\beta E_{n}}}{Z} \\
\times \left((\omega-\omega_{nm}+i\Gamma_{0}/2)^{-1}-(\omega+\omega_{nm}+i\Gamma_{0}/2)^{-1}\right),
\end{split}
\label{s1}
\end{equation}
where $|n\rangle$ and $E_{n}$ are the eigenstate and eigenvalue of the Hamiltonian, $\beta = 1/k_{B}T$, $Z = \sum_{n}e^{-\beta E_{n}}$ and $\omega_{nm} = (E_{n}-E_{m})/\hbar$. Here a phenomenological relaxation rate $\Gamma_{0}$ is introduced corresponding to the line width of the spectral shape. The conductivity is normalized with $\sigma_{0} = q^{2}Nx_{0}^{2}/\hbar$, where $x_{0}$ is a length scale, $x_{0}^2 = \hbar/\sqrt{Mk}$. The calculation indicates that the transition probability arises mostly from the matrix elements $\langle n+1|x|n \rangle$ between neighboring levels. Details are discussed in Ref.\,\cite{Matsumoto2009}.

\begin{figure}[h]
\includegraphics[width=9cm]{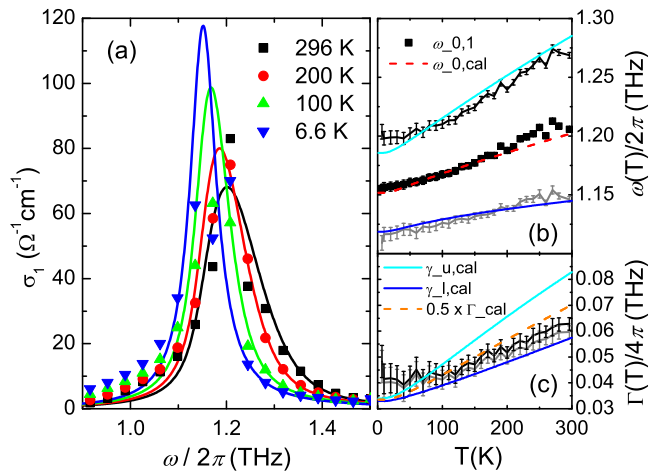}
\caption{(color online) (a) Spectral shapes of $\sigma_{1}(\omega)$ from $\nu_{1}$ of BGG. The filled symbols and solid lines indicate the data and calculations respectively. Temperature dependence of (b) the peak frequency $\omega_{0}$ and (c) the observed line widths compared with calculated ones. See the text for the definitions. }
\label{fig4}
\end{figure}

For the numerical calculation of $\sigma_{1}(\omega,T)$ for the $\nu_1$ mode, we fix the parameters: $(1/2\pi)\sqrt{k/M}$ =1.143\,THz, $\lambda/2k \times a_{B}^{2}=0.21$ ($a_{B}$ is Bohr radius), and the temperature-independent constant $\Gamma_{0}/2 \pi = 0.067\,\textrm{THz}$. With these parameters the observed spectra are best fitted to the calculations as shown in Fig.\,\ref{fig4}(a). Figure.\,\ref{fig4}(b) shows the temperature dependences of the peak position ($\omega_{0,cal}$) and the upper (lower) frequencies $\omega_{u(l),cal}$ giving half the peak height of the calculated spectra, together with corresponding data $\omega_{0,1}$, $\omega_{u,1}$ and $\omega_{l,1}$ obtained from the Lorentzian fitting. The data can be reproduced with our calculations.

In a usual harmonic approximation with $\lambda$ = 0, $E_{n} = \hbar\omega_{E}(n +1/2)$ and all the neighboring level-spacings become equivalent resulting in the well-known Lorentz-type dispersion in the complex conductivity for an harmonic oscillator $i\omega/(\omega^{2}-\omega_{E}^{2} + i\omega\Gamma_{0})$, as Eq.\ref{fitting}. In contrast, the anharmonicity due to the quartic term in Eq.\,(\ref{H}) induces unequally spaced energy levels, resulting in $n$-dependent excitations that the larger the $n$, the wider the spacing. At enough high temperatures in comparison to $\hbar\omega_{E}/k_B$ = 55\,K (= 1.15\,THz), phonons are thermally distributed widely from $n = 0$ to higher levels in accordance to the Boltzmann factor, while, at lower temperatures, phonons become condensed mostly around $n$ = 0, and can be effectively excited to $n$ = 1. The overall spectral shapes, given by a superposition of each Lorentzian curves having the $n$-dependent peak frequencies, become broad at higher temperature. Thus the anharmonicity effects on the conductivity spectra are expected both in the shift of the peak frequency and in the change of the line width.

Thus the 1D-AP model predicts an asymmetric change in the line width with temperature, which can be featured in the temperature dependence of $\gamma_{u(l),cal}= |\omega_{u(l),cal}- \omega_{0,cal}|$ measuring the width in the higher (lower)-frequency region with respect to the peak frequency. Figure\,\ref{fig4}(c) shows stronger temperature dependence in $\gamma_{u,cal}$ than in $\gamma_{l,cal}$, indicating the asymmetric change of the spectra. The data are well reproduced by the mean width defined as half the $\Gamma_{cal}=\gamma_{u,cal}+\gamma_{l,cal}$ shown by the dashed line.  
 
Finally, we remark upon two assumptions made in the above discussions. First, we have assumed the constant line-width ($\Gamma_{0}$) in conductivity calculations. This damping effect may be attributed, in part at least, to some mode-mode couplings to the anharmonic phonons themselves and also to the acoustic cage phonons. Second, we have assumed the positive quadratic potential ($k > 0$ in Eq.\,\ref{H}) meaning the on-center rattling in a single minimum potential. For the case of the off-centering with two-level minima with $k < 0$, the peak frequency and the line width are hardly fitted to the data. 

In conclusion, the temperature dependences of the peak frequency and the line width in the conductivity spectra of BGG are explained by 1D-AP model calculations without any interactions taken into account. Importantly to note, these dependences are determined solely by the Bose-factor for optical excitations of anharmonic phonons with the nonequally spaced energy levels.
 
The works have been supported by Grants-in-Aid for Scientific Research (A)(15201019,1820432), the priority areas (1951011, 15072205, 20102004) from MEXT, Japan, the Sasakawa Scientific Research from Japan Science Society, and also by the Global COE program ``Materials Integrations", Tohoku University.

\end{document}